\documentclass[runningheads]{llncs}
\usepackage{graphicx}
\usepackage{multirow}
\usepackage{amsmath}
\usepackage{amssymb}
\usepackage{biblatex}  
\usepackage{xcolor}
\begin{document}
\title{AID-DTI: Accelerating High-fidelity Diffusion Tensor Imaging with Detail-preserving Model-based Deep Learning}

\titlerunning{AID-DTI}

\author{Wenxin Fan\inst{1, 2} \and
Jian Cheng\inst{3} \and
Cheng Li\inst{1} \and
Jing Yang\inst{1, 2} \and
Ruoyou Wu\inst{1, 2, 4} \and
Juan Zou\inst{5} \and
Shanshan Wang\inst{1, 2, 4} \\
}
\authorrunning{W. Fan et al.}
%
\institute{
Paul C. Lauterbur Research Center for Biomedical Imaging, Shenzhen Institutes of Advanced Technology, Chinese Academy of Sciences, Shenzhen 518055, China \and
University of Chinese Academy of Sciences, Beijing 100049, China \and
Beihang University, Beijing 100191, China \and
Peng Cheng Laboratory, Shenzhen 518055, China \and
Changsha University of Science and Technology, School of Physics and Electronic Science, Changsha 410114, China \\
}

\maketitle              

\begin{abstract}
Deep learning has shown great potential in accelerating diffusion tensor imaging (DTI). Nevertheless, existing methods tend to suffer from Rician noise and eddy current, leading to detail loss in reconstructing the DTI-derived parametric maps especially when sparsely sampled q-space data are used. To address this, this paper proposes a novel method, AID-DTI (\textbf{A}ccelerating h\textbf{I}gh fi\textbf{D}elity \textbf{D}iffusion \textbf{T}ensor \textbf{I}maging), to facilitate fast and accurate DTI with only six measurements. AID-DTI is equipped with a newly designed Singular Value Decomposition-based regularizer, which can effectively capture fine details while suppressing noise during network training by exploiting the correlation across DTI-derived parameters. Additionally, we introduce a Nesterov-based adaptive learning algorithm that optimizes the regularization parameter dynamically to enhance the performance. AID-DTI is an extendable framework capable of incorporating flexible network architecture. Experimental results on Human Connectome Project (HCP) data consistently demonstrate that the proposed method estimates DTI parameter maps with fine-grained details and outperforms other state-of-the-art methods both quantitatively and qualitatively.

\keywords{diffusion tensor imaging  \and deep learning \and SVD.}
\end{abstract}
\section{Introduction}
Diffusion magnetic resonance imaging (dMRI) is a prominent non-invasive neuroimaging technique for measuring tissue microstructure. Among various dMRI techniques, diffusion tensor imaging (DTI) \cite{basser1994mr} is widely used to extract brain tissue properties and identify white matter tracts in vivo. The metrics from DTI, such as fractional anisotropy (FA), mean diffusivity (MD), and axial diffusivity (AD) \cite{curran2016quantitative} have great specificity in mapping the microstructural changes caused by normal aging \cite{salat2005age}, neurodegeneration \cite{thompson2013effectiveness}, and psychiatric disorders \cite{zheng2014dti}.  

To increase the accuracy of DTI-derived parametric maps, studies typically need more than the minimum of 6 diffusion weighting (DW) directions or acquire repeated observations of the same set of DW directions \cite{landman2007effects}. Moreover, {the low signal-to-noise ratio (SNR) poses significant challenges to subsequent analysis}, which further increases the demand for data to enable high-fidelity DTI metrics. Therefore, there is an urgent need to develop high-quality DTI metrics estimation from sparsely sampled q-space data.

Recently, deep learning has emerged as a powerful tool for accelerating DTI imaging. The pioneering work, q-space deep learning (q-DL) \cite{golkov2016q}, was introduced to directly map a subset of diffusion signals to Diffusion Kurtosis Imaging (DKI) parameters using a three-layer multilayer perceptron (MLP). Gibbons et al. \cite{gibbons2019simultaneous} used a 2D convolutional neural network (CNN) to estimate the Neurite Orientation Dispersion and Density Imaging (NODDI) and generalized fractional anisotropy maps. Similarly, SuperDTI \cite{li2021superdti} used deep CNN to model the nonlinear relationship between the acquired DWIs and the desired DTI-derived maps. In addition to data-driven mapping approaches, there has been a growing interest in model-driven neural networks that leverage domain knowledge to enhance network performance and interpretability. A notable example is the works proposed by Ye et al. \cite{ye2019deep, ye2020improved, zheng2023microstructure} which unfold the iterative optimization process for parameter mappings. Chen et al. used a subset q-space to estimate the parameters by explicitly considering the q-space geometric structure with a graph neural network (GNN) \cite{chen2020estimating, chen2022hybrid}. Furthermore, some excellent works in DWI super-angular-resolution can assist in the prediction of high-quality DTI metrics\cite{tian2020deepdti, chen2023super, tang2023high}.

Despite the progress made, the current methods still suffer from noise corruption or fine detail loss at a highly accelerated imaging rate. In this study, we propose a novel model-based deep learning method, named AID-DTI (\textbf{A}ccelerating h\textbf{I}gh fi\textbf{D}elity \textbf{D}iffusion \textbf{T}ensor \textbf{I}maging) to facilitate fast and accurate DTI. The main contributions of this work can be summarized as follows:
\begin{enumerate}
\item We propose a simple but effective model-based deep learning model, with a newly designed regularization to facilitate high-fidelity DTI metrics derivation. This term leverages the correlations and data redundancy between metrics, specifically targeting the alignment between predicted parameters and ground truth in singular-value subspaces, thus effectively capturing fine-grained details while suppressing noise.
\item We propose a novel Nesterov-based hyperparameter adaptive learning algorithm that integrates approximate second-order derivative information into the network training process, enabling more efficient hyperparameter tuning and better performance.
\item AID-DTI enables fast and high-fidelity DTI metrics estimation using a minimum of six measurements along uniform diffusion-encoding directions. Experiments demonstrated that our method outperforms current state-of-the-art methods both quantitatively and qualitatively.
\end{enumerate}

\section{Methods}
\begin{figure}
\includegraphics[width=\textwidth]{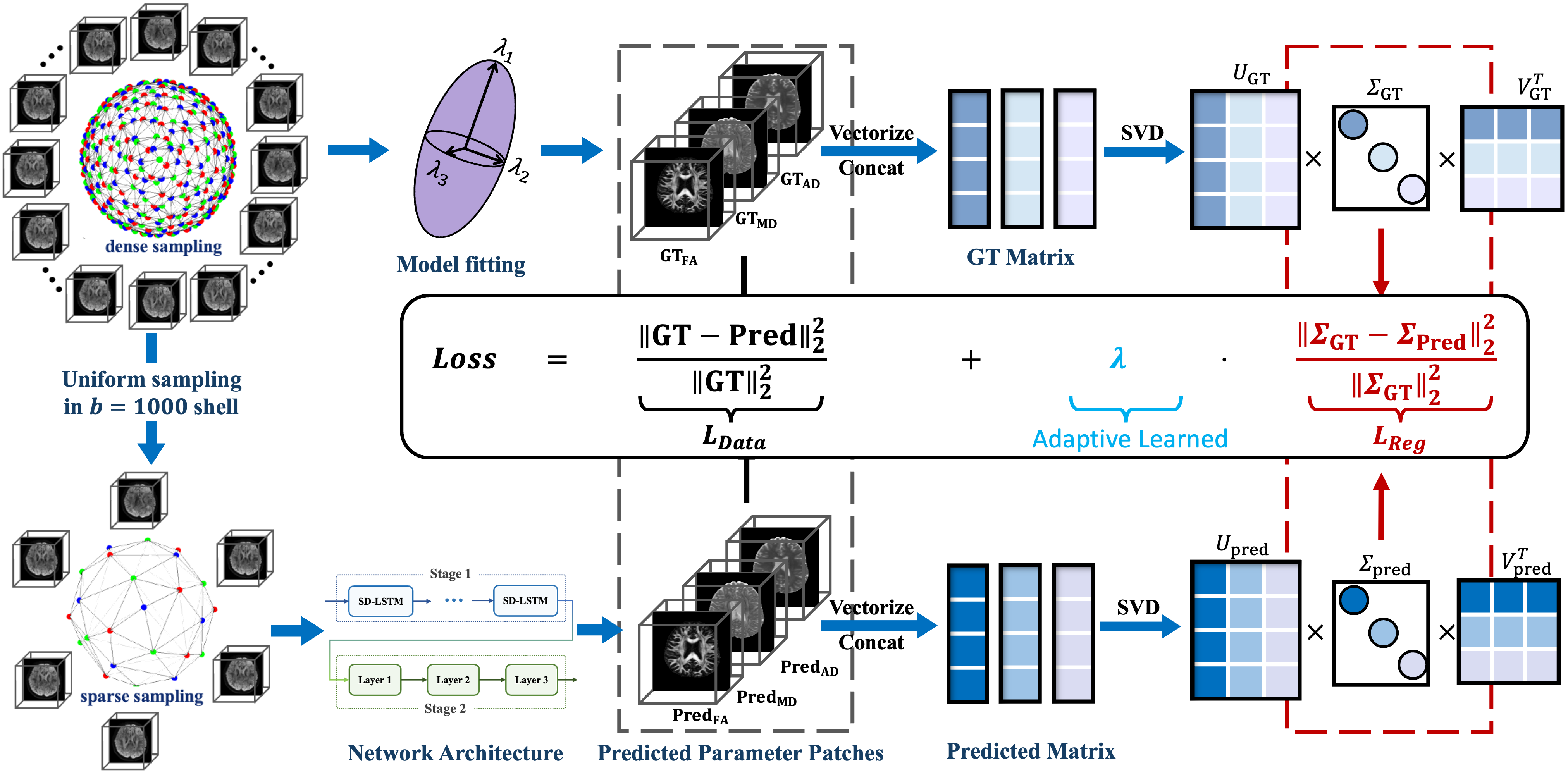}
\caption{{The proposed AID-DTI pipeline. The network input is super sparse measurements uniformly sampled from the dense measurements, and then the mapping between the sparsely sampled signal and three DTI metrics is directly learned simultaneously. After the network output, we vectorize each parameter and concatenate them into a new matrix, then perform SVD on this matrix to obtain the singular values. The weighted parameter $\lambda$ is adaptively learned to balance between data fidelity and SVD-regularization.}
} \label{fig1}
\end{figure}

In this section, we present the statement of the problem and a detailed presentation of the proposed AID-DTI, which investigates the accurate DTI-derived metric estimation using only six measurements instead of the recommended 30 measurements \cite{jones2013white} to achieve reliable prediction within the needed clinical accuracy. The proposed method is depicted in Fig. \ref{fig1} and encompasses key elements, specifically the SVD-based regularization (SVD-Reg) and the Nesterov-based adaptive learning algorithm (NALA).

As illustrated in Fig. \ref{fig1}, the overall architecture consists of two branches, with the upper branch representing the ground truth acquisition from dense sampling, while the lower branch symbolizes the network prediction from the sparse sampling. The network input is super sparse measurements uniformly sampled from the dense measurements using DMRITool \cite{cheng2015novel, cheng2017single}, and Singular Value Decomposition (SVD) is applied to both network prediction and ground truth to ensure the singular value consistency. 

\subsection{Task Formulation}
Our goal is to estimate reliable and fine-grained DTI parametric maps using only six measurements. Each diffusion signal can be considered as a set of $ W\times H\times S$ size volumes captured in the q-space. Thus the dMRI data are 4D signals of size $\mathbb{R}^{W\times H\times S\times D}$, where $W$, $H$, $S$, $D$ refer to the width, height, slice, and gradient directions, respectively. 

Given the diffusion MRI data ${X} \in \mathbb{R}^ {W \times H \times S \times D_{Full}}$ containing the full measurements in the q-space, the ground-truth scalar maps $Y_{GT}$ obtained from all the diffusion data, we aim to design a network $\mathcal{F}_\theta$ parameterized by $\theta$ to learn a mapping from the given sparsely sampled signal $\tilde{X}{\in\mathbb{R}}^{W\times H\times S\times D_{Sparse}}$ to predicted DTI metrics ${Y}$, s.t $Y=\mathcal{F}_\theta(\tilde{X})\rightarrow Y_{GT}$.

\subsection{SVD-based Regularization}

Most existing regularization strategies only consider the properties of the diffusion signal and apply the regularization to the DWI data rather than the desired parametric maps, such as sparsity \cite{cheng2015joint, yap2016multi, schwab2018joint, schwab2019global}, low-rank \cite{cheng2015joint, shi2016super, zhang2020acceleration, ramos2021snr}, total variation \cite{liu2014generalized, shi2016super, teh2020improved} regularization, etc. To facilitate accurate and fine-grained DTI metric prediction, we explicitly consider the quality of derived parameters and propose the incorporation of an SVD-based regularization term to enhance performance. 

\begin{equation}
Loss = L_{Data} + \lambda \cdot R
=\frac{\|Y_{GT}-Y_{Pred}\|_2^2}{\|Y_{GT}\|_2^2} +\lambda \cdot \frac{\left\|\Sigma_{{GT}}-\Sigma_{{Pred}}\right\|_2^2}{\left\|\Sigma_{\mathrm{GT}}\right\|_2^2}
\end{equation}

{
The actual input in our implementation is the $N \times N \times N$ patches instead of the whole DWI volume, so the output of the network $Y_{\mathrm{Pred}}$ is $N \times N \times N \times 3$, the last dimension indicating three parameter maps. Then, we vectorize each parameter and concatenate them into a new matrix, referred to as the parameter matrix. We perform SVD on the predicted matrix and GT matrix respectively to obtain the singular values $\Sigma_{\mathrm{Pred}}$ and $\Sigma_{\mathrm{GT}}$. 
}

From the statistical point of view, the singular matrices of a data matrix represent the principal component directions, i.e., the directions that exhibit the highest variance corresponding to the largest singular values. According to the Eckart–Young theorem \cite{eckart1936approximation}, the dominant singular subspaces capture the majority of the informational content. {It can be believed that the major singular values encapsulate the dominant features of the three parameters. Therefore, ensuring the consistency of the primary singular values preserves the integrity of the extracted significant information,} effectively maintaining fine details while reducing a certain level of noise. {Subsequent denoising experiments also demonstrated the superiority of the proposed method in noise handling.
}

\subsection{Nesterov-based Hyperparameter Adaptive Learning Algorithm}
The total loss is the weighted combination of the data-fidelity term and the proposed regularization term. However, the process of hyperparameter selection is in practice often based on trial-and-error and grid or random search \cite{bengio2000gradient,duchi2011adaptive,maclaurin2015gradient,franceschi2017forward}, which can be a time-consuming process.


Building upon the foundation laid by previous studies\cite{almeida1999parameter, baydin2017online, rubio2017convergence, chandra2022gradient}, we propose a Nesterov-based hyperparameter adaption algorithm.
{The hyperparameter optimization problem is inherently a bilevel optimization task because of its hierarchical nature \cite{sinha2017review, franceschi2017forward, franceschi2018bilevel}. The outer problem requires minimizing the validation set loss, concerning the hyperparameter $\lambda$, and the inner problem requires minimizing the training set loss, for the model parameter $\theta$.} Thus, our method optimizes the network parameter $\theta$ and hyperparameter $\lambda$ alternately on the training and validation sets, respectively, {which means the $\lambda$ that minimizes the validation loss will be accepted}. Let $\lambda_t$ and $\theta_t$ be the values of $\lambda$ and $\theta$ at the step $t$. More specifically, the iterations go as follows:
\begin{equation}
\left\{
\begin{aligned}
    L(\theta,\lambda) &=L_{Data}(\theta)+\lambda \cdot R(\theta) \\
    \text{On Training Set: } \theta_{t+1} &= \arg \min_\theta{L(\theta,\lambda_t)} \\
    \text{On Validation Set: } \lambda_{t+1}&= \arg\ \min_\lambda{L(\theta_{t+1},\lambda)}
\label{iterative_update}
\end{aligned}
\right.
\end{equation}


In analogy to updating network parameter $ \theta$, $\lambda$ should be updated in the direction of the gradient of the $Loss\left(\theta,\lambda\right)$ concerning $\lambda$, scaled by another hyper-hyperparameter $\beta$. One way to compute $\frac{\partial L(\theta_{t+1},\lambda_{t})}{\partial\lambda_t}$ is the direct manual computation of the partial derivative:
\begin{equation}
\beta \cdot \frac{\partial Loss\left(\theta_{t+1},\lambda\right)}{\partial\lambda}=\beta \cdot \frac{\partial\left[L_{Data}\left(\theta_{t+1}\right)+\lambda \cdot R\left(\theta_{t+1}\right)\right]}{\partial\lambda}= \beta \cdot R(\theta_{t+1})
\label{gradient}
\end{equation}

In other words, the adjustment at step $t+1$ depends on the regularization term value. This expression lends itself to a simple and efficient implementation: simply remember the past regularization value. By leveraging insights from the Nesterov accelerated gradient (NAG) \cite{nesterov2013gradient,nesterov1983method}, which has a provably bound for convex, non-stochastic objectives, we introduce an improved momentum term $m$ here:
\begin{equation}
\left\{
\begin{aligned}
m_{t+1} &=\beta \cdot m_t + R(\theta_{t+1}) + \beta \cdot \left[R(\theta_{t+1})-R(\theta_{t})\right] \\
\lambda_{t+1} &=\lambda_t-\kappa \cdot m_{t+1} \\
\end{aligned}
\right.
\end{equation}
where $R(\theta_{t+1})-R(\theta_{t})$ is actually the differential of the gradient concerning $\lambda$, which approximates the second-order derivative of the objective function. Thus, the improved momentum term $m_{t+1}$ is the combination of the past search directions $m_t$, current stochastic gradient $R(\theta_{t+1})$, and the approximate second-order derivative $R(\theta_{t+1})-R(\theta_{t})$.

\subsection{Backbone Network}
Here, the Microstructure Estimation with Sparse Coding using Separable Dictionary (MESC-SD) \cite{ye2020improved}, an unfolding network based on sparse LSTM units \cite{zhou2018sc2net} with two cascaded stages, is employed. The first stage computes the spatial-angular sparse representation of the diffusion signal while the second stage maps the sparse representation to tissue microstructure estimates. Note that AID-DTI is highly versatile, such that networks producing output in a matrix or higher-dimensional tensor form are compatible with our methodology. We can support CNN \cite{gibbons2019simultaneous}, MESC-SD \cite{ye2020improved}, and even one-dimensional networks like q-DL \cite{golkov2016q} can benefit from our method by appropriately reshaping their outputs into matrix form.

\section{Experiments and Results}
\subsection{Dataset}
Pre-processed whole-brain diffusion MRI data from the publicly available Human Connectome Project (HCP) dataset were used for this study \cite{van2013wu}. Our dataset consists of 111 subjects randomly selected from the HCP, which is partitioned into 60 subjects for training, 17 subjects for validation, and 34 subjects for testing. 

To obtain the input data of AID-DTI, DWI volumes acquired along six uniform diffusion-encoding directions at $b=1000s/mm^{2}$ of each subject were selected using DMRITool \cite{cheng2015novel, cheng2017single}. To obtain the ground-truth DTI metrics, diffusion tensor fitting was performed on all the diffusion data using ordinary linear squares fitting implemented in the DIPY\footnote{https://github.com/dipy} software package to derive the FA, MD, and AD \cite{curran2016quantitative}. 

\subsection{Comparison Methods}
Our method was compared qualitatively and quantitatively with DIPY, which represents the conventional DTI model fitting (MF) algorithm, and deep learning-based approaches, including the q-DL in Golkov et al. [14], CNN in Gibbons et al. [15], MESC-SD in Ye et al. [19].

\subsection{Implementation Details}
The neural network was implemented using the PyTorch library (codes will be available online upon acceptance of the paper). We trained the network with Two Tesla V100 GPUs (NVIDIA, Santa Clara, CA) with 32GB memory. All networks adopted the Adam optimizer and the learning rates were initialized as 0.01, 0.001, and 0.0001 respectively. For all the networks, the extracted brain masks from the preprocessing pipeline were applied to only include voxels within the brain when evaluating the performance.

\subsection{Evaluation Results}
We evaluate the performance of AID-DTI through a comparative analysis with baseline methods and other state-of-the-art methods by computing the SSIM and PSNR to quantify the similarity compared to the ground truth. The experimental results are summarized in Table \ref{tab1}, where it can be observed that AID-DTI surpasses the comparison methods by a large margin.


\begin{table}
\begin{center}
\caption{{The quantitative results were obtained with 6 diffusion directions at b-values of 1000 $s/{mm}^2$ in terms of MSE, SSIM, and PSNR. The best results are in \textbf{bold}.}}\label{tab1}
\resizebox{\textwidth}{!}{
\begin{tabular}{|c|c c c c|c c c c|c c c c|}
\hline
\multirow{2}{*}{Methods} & \multicolumn{4}{c|} {MSE $\times 10^{-3}$}  & \multicolumn{4}{c|} {SSIM} & \multicolumn{4}{c|} {PSNR}\\
\cline{2-13}
& FA & MD & AD & All& FA & MD & AD & All & FA & MD & AD & All \\
\cline{1-13}
\hline
MF &36.00 &15.55 &26.38 &25.98$\pm$3.74  &0.719 &0.760 &0.714 &0.772$\pm$0.031 &14.437 &18.082 &15.786 &15.854$\pm$0.618 \\
q-DL &2.293 &1.160 &1.419 &1.623$\pm$0.23  &0.904& 0.952 &0.931 &0.929$\pm$0.009 &26.397 &29.354 &28.481 &27.894$\pm$\textbf{0.591} \\
CNN &1.184 &0.726 &0.934 &0.948$\pm$0.16  &0.941 &0.968 &0.951 &0.953$\pm$0.007 &29.266 &31.390 &30.299 &30.232$\pm$0.679 \\
MESC-SD &0.756 &0.671 &0.824 &0.750$\pm$0.13  &0.952 &0.971 &0.958 &0.960$\pm$0.006 &31.216 &31.733 &30.841 &31.248$\pm$0.689 \\
Ours &\textbf{0.683} &\textbf{0.626} &\textbf{0.774} &\textbf{0.694$\pm$0.12}  &\textbf{0.956} &\textbf{0.973} &\textbf{0.961} &\textbf{0.963$\pm$0.005} &\textbf{31.653} &\textbf{32.037} &\textbf{31.113} &\textbf{31.638}$\pm$0.676 \\

\hline
\end{tabular}}
\end{center}
\end{table}

\begin{figure}
\centering
\includegraphics[width=0.7\textwidth]{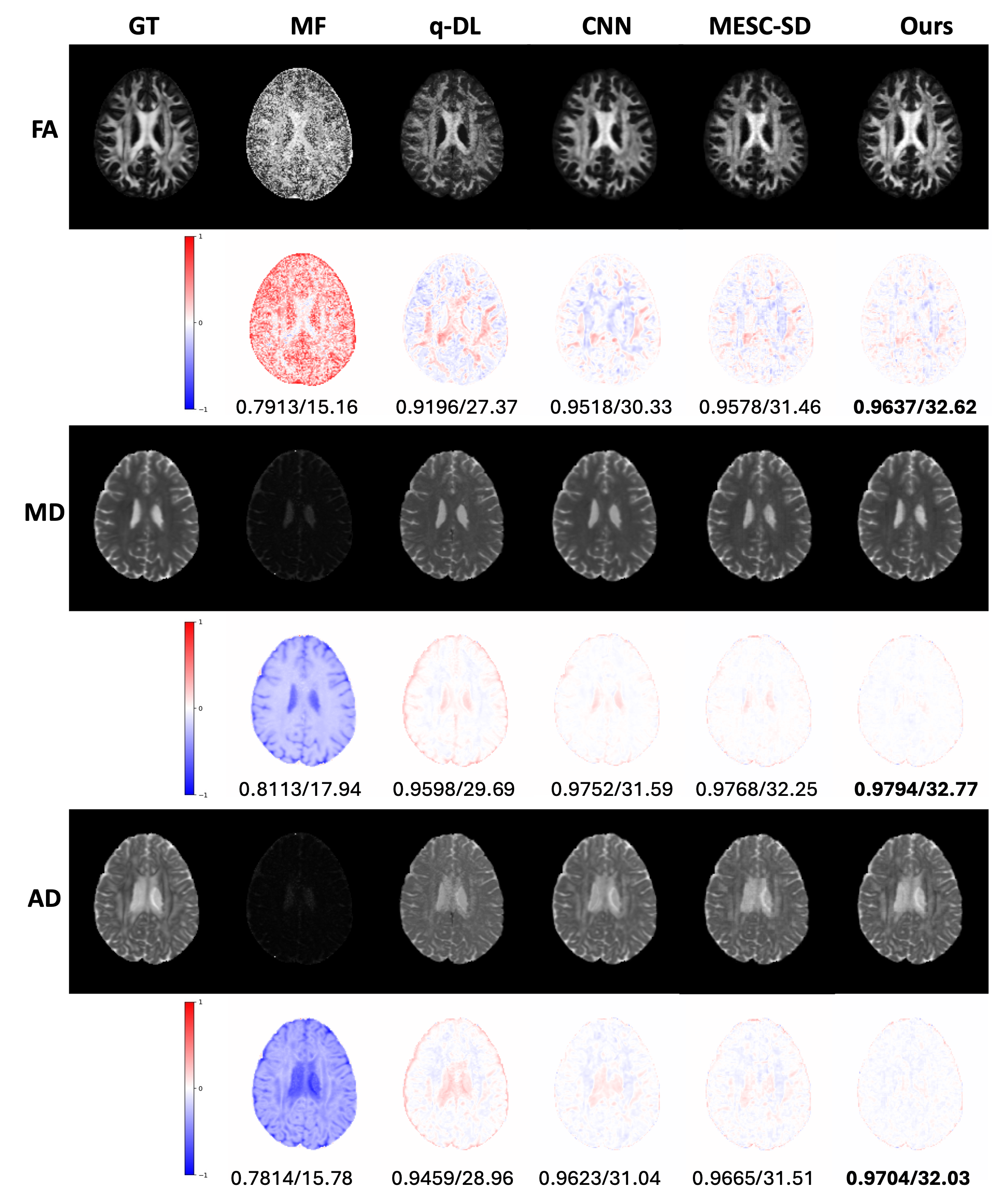}
\caption{The ground truth, estimated DTI parameters FA, AD, and MD, and corresponding residual maps based on MF, q-DL, CNN, MESC-SD (baseline), and Ours in a test subject with 6 diffusion directions at b-values of 1000$s/{mm}^2$.} \label{fig2}
\end{figure}

For qualitative analysis, we provide the estimation results in Fig. \ref{fig2}. As can be seen, the conventional method MF produces significant estimation error and loses anatomical information when only six measurements were employed. The results from the figure also show that the q-DL method yields a relatively low signal-to-noise ratio, while the CNN method, although achieving better results, appears overly smooth in qualitative images, leading to the loss of texture. MESC-SD, as one of the state-of-the-art microstructure estimation methods, showcases excellent results, when our method combined with it, demonstrates enhanced performance as evidenced by the error maps, effectively preserving crucial anatomical details.

\begin{figure}
\centering
\includegraphics[width=0.75\textwidth]{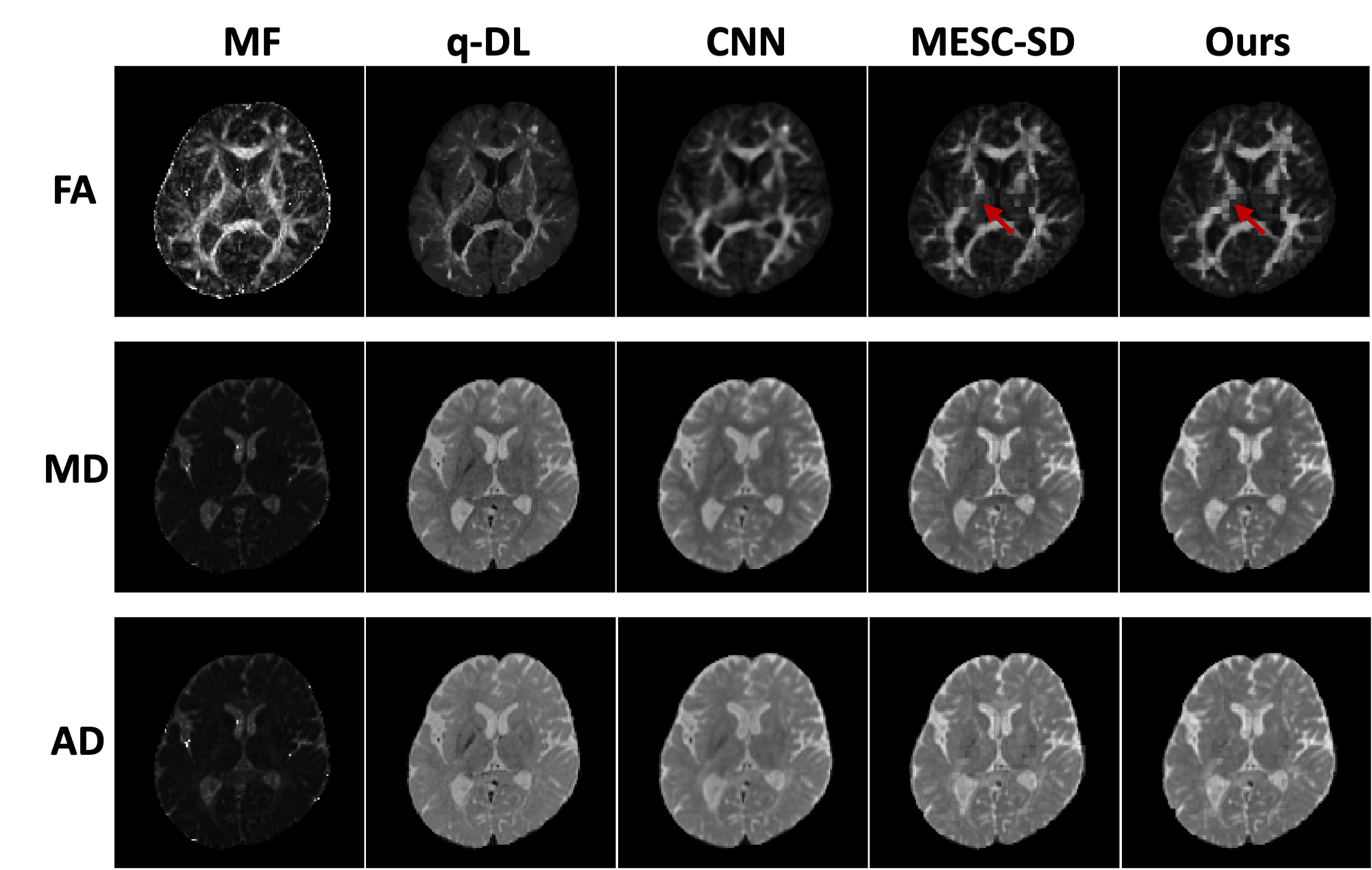}
\caption{{Prospective results in a test subject with real low angular resolution data (6 diffusion directions at b-values of 1000$s/{mm}^2$ and 2 at $b_0$).}} \label{fig3}
\end{figure}

To validate the noise-handling capabilities of the proposed method, we synthesize noisy data by introducing Rician noise at levels of $2.5\%$ and $5\%$ into the diffusion-weighted signals. Then input the noise-corrupted data into the trained networks to predict the three DTI-derived scalar maps. Table \ref{tab2} shows the comparative results for varying noise levels. To ensure a fair comparison, we also considered applying denoising algorithms after MF, specifically BM4D\footnote{https://pypi.org/project/bm4d/}\cite{dabov2007image, maggioni2012nonlocal} was chosen in our experiments. Both Table \ref{tab1} and Table \ref{tab2} demonstrate that our method outperforms others in clean and noisy conditions, indicating a degree of noise resistance.

\begin{table}
\centering
\caption{{Quantitative evaluation of denoising performance using synthetic data with different level of Rician noise. The best results are in \textbf{bold}.}}\label{tab2}
\resizebox{\textwidth}{!}{
\begin{tabular}{|c|c c c |c c c |}
\hline
\multirow{2}{*}{Methods} & \multicolumn{3}{c|} {$\sigma=0.025$}  & \multicolumn{3}{c|} {$\sigma=0.05$}\\
\cline{2-7}
& MSE $\times 10^{-3}$ & SSIM & PSNR & MSE $\times 10^{-3}$ & SSIM & PSNR \\
\cline{1-7}
\hline
MF &28.32$\pm$3.57 &0.760$\pm$0.028 &15.511$\pm$\textbf{0.538} &28.67$\pm$3.58 &0.759$\pm$0.029 &15.425$\pm$\textbf{0.540} \\
MF+BM4D &28.07$\pm$3.56  &0.761$\pm$0.029 &15.551$\pm$0.542 &28.59$\pm$3.58  &0.759$\pm$0.029 &15.437$\pm$0.542 \\
q-DL &1.604$\pm$0.27 &0.914$\pm$0.012 &27.290$\pm$0.600  &2.560$\pm$0.38 &0.882$\pm$0.016 &25.858$\pm$0.619\\
CNN &1.111$\pm$0.17  &0.942$\pm$0.008 &29.545$\pm$0.652  &1.609$\pm$\textbf{0.25}  &0.917$\pm$\textbf{0.011} &27.934$\pm$0.663\\
MESC-SD &0.941$\pm$0.16  &0.947$\pm$0.008 &30.266$\pm$0.682 &1.586$\pm$\textbf{0.25}  &0.922$\pm$0.012 &28.000$\pm$0.655\\
Ours &\textbf{0.893$\pm$0.15}  &\textbf{0.952$\pm$0.007} &\textbf{30.547$\pm$}0.680 &\textbf{1.499}$\pm$0.27  &\textbf{0.927}$\pm$0.012 &\textbf{28.302$\pm$}0.732\\
\hline
\end{tabular}}
\end{table}

Due to the absence of publicly available super low angular resolution (six-direction) datasets, we used in-house data here to conduct prospective experiments. The imaging protocol was as follows: 2 $b0$ gradient directions and 6 b=1000$s/mm^2$ gradient directions; 140 × 140 imaging matrix; voxel size $1.5 \times 1.5 \times  1.0 mm^3$; TE/TR = 66.0/5,820 ms.

\subsection{Ablation Study}

In this section, we perform an extensive ablation study to investigate the effectiveness of the SVD-based regularization (SVD-Reg) module and Nesterov-based adaptive learning algorithm (NALA). As shown in Table \ref{tab3}, the ablation study is completed under the condition of 6 gradients at a b-value of 1000 $s/{mm}^2$. Table \ref{tab2} shows the quantitative results of the three variants, respectively. According to the quantitative results, the average values of PSNR and SSIM achieved by AID-DTI are the highest among the three variants.


\begin{table}
\centering
\caption{Ablation results using {MSE}, SSIM, and PSNR. The best results are in \textbf{bold}.}\label{tab3}
\begin{tabular}{|c|c|c|c|c|c|}
\hline
{Models} & {SVD-Reg} & {NALA} & {MSE ($\times 10^{-3}$)} & {SSIM} & {PSNR} \\
\cline{1-6}
\hline
(A) & & &0.750$\pm$0.13 &0.960$\pm$0.006  &31.248$\pm$0.689\\
(B) &\checkmark & &0.704$\pm$0.12  &0.962$\pm$0.005 &31.525$\pm$0.690 \\
(C) &\checkmark &\checkmark &\textbf{0.694$\pm$0.12} &\textbf{0.963$\pm$0.005}  &\textbf{31.638$\pm$0.676}\\
\hline
\end{tabular}
\end{table}



\section{Conclusion}
In this study, we develop a novel model-driven deep learning approach AID-DTI for reducing the q-space sampling requirement of DTI. Our method maps one b=0 image and six DWI volumes to high-quality DTI metrics employing an SVD-based regularization and introduces an adaptive algorithm for automatically updating regularization parameters. The proposed method exhibits simplicity, flexibility and has a high potential to become a practical tool in a wide range of clinical and neuroscientific applications. Future efforts will expand the proposed method to other diffusion models and more multi-parametric MR imaging scenarios.

\subsubsection{Acknowledgements}
This research was partly supported by the National Natural Science Foundation of China (62222118, U22A2040), Guangdong Provincial Key Laboratory of Artificial Intelligence in Medical Image Analysis and Application (2022B1212010011), Shenzhen Science and Technology Program (RCYX20210706092104034, JCYJ20220531100213029), and Key Laboratory for Magnetic Resonance and Multimodality Imaging of Guangdong Province (2023B1212060052).
%
%
%
\printbibliography

\end{document}